\documentclass[fleqn,preprint,showpacs,preprintnumbers,amsmath,amssymb]{revtex4}
\usepackage{graphicx}
\usepackage{dcolumn}
\usepackage{bm}

\begin{document}

\title{Orbital Ferromagnetism and Quantum Collapse in Stellar Plasmas}
\author{M. Akbari-Moghanjoughi}
\affiliation{Azarbaijan University of
Tarbiat Moallem, Faculty of Sciences,
Department of Physics, 51745-406, Tabriz, Iran}

\date{\today}
\begin{abstract}
The possibility of quantum collapse and characteristics of nonlinear localized excitations is examined in dense stars with Landau orbital ferromagnetism in the framework of conventional quantum magnetohydrodynamics (QMHD) model including Bohm force and spin-orbit polarization effects. Employ the concepts of effective potential and Sagdeev pseudopotential, it is confirmed that the quantum collapse and Landau orbital ferromagnetism concepts are consistent with the magnetic field and mass-density range present in some white dwarf stars. Furthermore, the value of ferromagnetic-field found in this work is about the same order of magnitude as the values calculated earlier. It is revealed that the magnetosonic nonlinear propagations can behave much differently in the two distinct non-relativistic and relativistic degeneracy regimes in a ferromagnetic dense astrophysical object. Current findings should help to understand the origin of the most important mechanisms such as gravitational collapse and the high magnetic field present in many compact stars.
\end{abstract}

\maketitle

\section{Introduction}

One of the greatest challenges in astrophysics is to understand the mysterious origin of high magnetic field and unique physical processes in many compact stars. There has been spectacular breakthroughs in this area of science in the past several years since the pioneering works of Chandrasekhar, Bohm, Canuto, Chiu and O'Connell et.al \cite{chandra1, bohm, can1, con1, weeler} on relativistic degeneracy and magnetism. We have already observed manu peculiar properties that arise due to Pauli exclusion and electron degeneracy in ordinary metallic and semiconductor materials such as quantum tunneling, Landau quantization, De Haas-van Alphen effect and many others \cite{haug, landau}. However, due to known experimental restrictions in astrophysical sciences, unfortunately, we are left only with observational and theoretical tools. Initial theoretical studies reveal that the relativistic degeneracy \cite{chandra2, chandra3} and magnetism \cite{can2, can3, can4, can5, can6, can7, can8} can lead to quite different equations of state and yet unknown phenomena such as quantum collapse \cite{chai, akbari1} and Landau orbital ferromagnetism (LOFER) \cite{con2, burk1}. A recent complete review of different aspects of a dense super-magnetized matter may be found in Refs. \cite{dong, harding}. Adam \cite{adam}, using various magnetic models including LOFER, has investigated the hydrostatic equilibrium and collapse phenomenon in a white dwarf star and compared the theory with observations. Based on the theoretical calculation of Lee et.al \cite{lee}, O'Connell \cite{con3} has found a stable solution for LOFER sustained self-field which might contribute to origin of compact astrophysical objects. O'Connell's initial calculation show inconsistent results \cite{burk2} for the temperatures observed in typical white dwarfs, however, it has been shown that the inclusion of electron exchange energy can lead to partial agreement but with slightly higher magnetic fields than measured in those objects.

On the other hand, more recently, Brodin and Marklund \cite{brodin0} have extended the quantum magnetohydrodynamics (QMHD) model to include the spin dipole effects and evaluated the dynamical behavior of nonlinear waves in an electron-ion plasma. They have reported an extreme regimes in which a spin-1/2 quantum plasma exhibits ferromagnetic-type properties. In recent years there has been great majority of activities in the field of both quantum hydrodynamics (QHD) and quantum magnetohydrodynamics (QMHD) \cite{haas1, haas2, akbari2, akbari3, Markowich, Marklund1, Brodin1, Marklund2, Brodin2, manfredi, shukla, gardner} which reveal the unexpected nonlinear dynamical features in quantum plasmas. Recently, using a classical Thomas-Fermi approximation \cite{akbari4, akbari5} and fully QMHD model \cite{akbari6} it has been confirmed that the nonlinear wave properties of a relativistically degenerate plasma fundamentally differs from those of an ordinary degenerate one. Also, other recent studies, employing the spin-included QMHD model justify the dominant spin effects on nonlinear wave properties in quantum plasmas \cite{marklund3, marklund4, misra1, martin, mushtaq, brodin3, zaman, vitaly, misra2}. In this paper, we employ the spin-included quantum magnetohydrodynamics model to explore the possibility of Landau orbital ferromagnetism and quantum collapse in astrophysical dense objects such as white dwarfs and their possible role on nonlinear wave dynamics. The presentation of the paper is as follows. The LOFER QMHD model including Bohm potential and the spin-orbit magnetization effects is introduced in Sec. \ref{equations} and the Localized LOFER density excitations are evaluated and the appropriate pseudopotential is found in Sec. \ref{calculation}. The numerical results and the comparison of findings with the previously reported values are presented in Sec. \ref{discussion}. Finally, a summary is given in Sec. \ref{conclusion}.

\section{LOFER QMHD Model}\label{equations}

Considering a completely degenerate electron-ion plasma, including the electron spin-orbit polarization, quantum tunneling and Lorentz force effects, one may setup the extended QMHD equations. For instance, in a center of mass frame of plasma, the continuity equation reads as
\begin{equation}\label{cont}
\frac{{\partial \rho }}{{\partial t}} + \nabla \cdot(\rho {\bf{u}}) = 0,
\end{equation}
and, the extended momentum equation in cgs units is given as \cite{brodin0}
\begin{equation}\label{mom}
\rho \left[ {\frac{\partial }{{\partial t}} + {\bf{u}} \cdot \nabla } \right]{\bf{u}} =  - \nabla \left[ {\frac{{{B^2}}}{{2{}}} - {\bf{M}} \cdot {\bf{B}}} \right] - \nabla P + {\bf{B}} \cdot \nabla \left[ {{{\bf{B}}}{{{}}} - {\bf{M}}} \right] + \frac{{\rho {\hbar ^2}}}{{2{m_e}{m_i}}}\nabla \frac{{\Delta \sqrt \rho  }}{{\sqrt \rho  }},
\end{equation}
where the field due to the physical current, $\bf{H}$, magnetization, $\bf{M}$ and the sustained field, $\bf{B}$, are related through the relation; $\bf{H}=\bf{B}-4\pi \bf{M}$. Moreover, the ferromagnetic state of plasma is defined by setting $\bf{H}=0$ (i.e. $\bf{B}=4\pi \bf{M(B)}$) \cite{sho}. Canuto et.al have reported a stable LOFER state in strongly magnetized degenerate electron-gas leading to a magnetic equation of state of form $B = \alpha(\rho_6/\mu_e)^{2/3}$ \cite{con1, burk2, adam}, where, $\rho_6=\rho/10^6$ and $\mu_e$ is the number of nucleons per electron (for instance, $\mu_e=2$ for helium atom). The value of proportionality constant $\alpha$ is not a priori-evident \cite{adam}, however, it is known to be sensitively dependent on many parameters such as the plasma temperature, electron exchange interactions and the maximum magnetic-field oscillation amplitude \cite{burk1}. It is also noted that the value of $\alpha$ can change due to rapid oscillations in magnetic moment. In what follows, it will be clearly evident that our approach and conclusions are independent of the value of this parameter. Therefore, the QMHD equations can be closed using a magnetic equation of state such as that mentioned above.

On the other hand, the pressure of a relativistically degenerate magnetized Fermi-gas, i.e. $P_d$, (for $B /B_c \ll 1$, with $B_c = 4.414 \times 10^{13}G$) is given by \cite{chandra1}
\begin{equation}\label{p}
P\simeq{P_d(R)} = \frac{{\pi m_e^4{c^5}}}{{3{h^3}}}\left[ {R\left( {2{R^2} - 3} \right)\sqrt {1 + {R^2}}  + 3{{\sinh}^{ - 1}}R} \right],
\end{equation}
where, the Chandrasekhar relativity parameter, $R=P_{Fe}/(m_e c)=(\rho/\rho_c)^{1/3}$ ($\rho_c\simeq 2\times 10^6 gr/cm^3$) is a measure of the relativistic degeneracy of electrons and $P_{Fe}$ is the electron Fermi relativistic-momentum. It is noted that, in the extreme limits of very small relativity parameter, $R\ll R_{Ch}$ (the value $R_{Ch}=\sqrt{2}$ being the Chandrasekhar critical value of the relativity parameter), a dependence $P_d \propto \rho^{5/3}$ and for corresponding very large values, $R\gg R_{Ch}$, a dependence $P_{d}\propto \rho^{4/3}$ is expected for the electron degeneracy pressure.

For simplicity and without loss of the generality of the problem, we may assume that the propagation of magnetoacoustic nonlinear waves takes place along the $x$ direction perpendicular to the high induced magnetic-field $B(x,t)$ along the $z$-axis. Then, we derive
\begin{equation}\label{dimensional}
\begin{array}{l}
\frac{{\partial \rho }}{{\partial t}} + \frac{{\partial \rho u}}{{\partial x}} = 0, \\
\frac{{\partial u}}{{\partial t}} + u\frac{{\partial u}}{{\partial x}} =  - \frac{{{m_e}{c^2}}}{{{m_i}}}\frac{\partial }{{\partial x}}\left[ {\int {\frac{{{d_R}{P_d}(R)}}{R}} dR} \right] + \frac{{{m_e}{c^2}}}{{{m_i}}}\frac{{{\beta ^2}}}{{2\pi }}\frac{{\partial {R}}}{{\partial x}} + \frac{{{\hbar ^2}}}{{2{m_e}{m_i}}}\frac{\partial }{{\partial x}}\left[ {\frac{1}{{\sqrt \rho  }}\frac{{{\partial ^2}\sqrt \rho  }}{{\partial {x^2}}}} \right]. \\
\end{array}
\end{equation}
Note that we have defined a new parameter $\beta^2  = \alpha^2 ({m_i}/{m_e}{c^2})$. We also define the effective quantum potential, the so-called total quantum potential acting on fermions, as
\begin{equation}\label{dimensional}
{\Psi _{eq}} = \frac{{{\beta ^2}R}}{{2\pi }} - \int {\frac{{{d_R}{P_d}(R)}}{R}} dR = \frac{{{\beta ^2}R}}{{2\pi }} - \sqrt {1 + {R^2}} . \\
\end{equation}
The dimensionless set of QMHD equations is thus obtained using the following standard scalings
\begin{equation}\label{normal}
x \to \frac{{{c_{s}}}}{{{\omega _{pi}}}}\bar x,\hspace{3mm}t \to \frac{{\bar t}}{{{\omega _{pi}}}},\hspace{3mm}\rho \to \bar \rho{\rho_0},\hspace{3mm}u \to \bar u{c_{s}},
\end{equation}
where the normalizing factors, $\rho_0$, ${\omega _{pi}} = \sqrt {{e^2}{\rho_{0}}/(\varepsilon_0{m_i^{2}})}$ and ${c_{s}} = c\sqrt {{m_e}/{m_i}}$ denote the equilibrium plasma mass-density, characteristic ion plasma frequency and ion quantum sound-speed, respectively. After trivial simplifications, we arrive at the following equations
\begin{equation}\label{dimensionless}
\begin{array}{l}
\frac{{\partial \bar \rho}}{{\partial t}} + \frac{{\partial \bar \rho{\bar u}}}{{\partial x}} = 0, \\
\frac{{\partial {\bar u}}}{{\partial t}} + {\bar u}\frac{{\partial {\bar u}}}{{\partial x}} = \frac{{\partial {\Psi _{eq}}}}{{\partial x}} + {H^2}\frac{\partial }{{\partial x}}\left( {\frac{1}{{\sqrt {\bar \rho} }}\frac{{{\partial ^2}\sqrt {\bar \rho} }}{{\partial {x^2}}}} \right), \\
{\Psi _{eq}} = \frac{{{\beta ^2\bar\rho^{1/3}R_0}}}{{2\pi }}  - \sqrt {1 + {\bar\rho^{2/3}R_0^2}}. \\
\end{array}
\end{equation}
where, the bar notations denote the dimensionless quantities. We have also defined two other new parameters, namely, the relativistic degeneracy parameter $R_0=(\rho_0/\rho_c)^{1/3}$ and the quantum diffraction parameter, $H = \sqrt {{m_i}/{2m_e}}(\hbar {\omega _{pi}})/({m_e}{c^2})$ interrelated through $H = e\hbar \sqrt {{\rho_c}R_0^3/m_i\pi } /(2m_e^{3/2}{c^2})$. The normality notations here and after are avoided for clarity. It is realized that the gradient of the effective quantum potential defines the local quantum force acting on electrons which is used to describe the quantum-collapse phenomenon and is therefore the sum of the quantum-forces due to relativistic electron degeneracy pressure and the relativistic spin-orbit magnetization. The quantum collapse \cite{akbari1} is defined as a state of a plasma for which the effective quantum potential vanishes.

\section{LOFER Density Excitations}\label{calculation}

Since, we are interested in studying the stationary solitary excitations, it is convenient to change into the co-moving frame by defining a new variable as $\xi=x-V t$, and define the Eqs. (\ref{dimensionless}) in new coordinate. The frame speed, $V=u/c_s$ is then the Mach-number of the nonlinear magnetoacoustic excitations. Integration of the continuity equation regarding the new variable, $\xi$, making use of the boundary conditions $\mathop {\lim }\limits_{\xi  \to  \pm \infty } \rho = 1$ and $\mathop {\lim }\limits_{\xi  \to  \pm \infty } u = 0$ leads to the relation, $u = V\left( {{1}/{\rho} - 1} \right)$. Using the value for, $u$, in the momentum equation, with $\rho=A^2$, and integrating again with the above-mentioned boundary conditions, reduces the Eqs. (\ref{dimensionless}) to the following simplified nonlinear differential equation
\begin{equation}\label{diff2}
\frac{{{H^2}}}{A}\frac{{{\partial ^2}A}}{{\partial \xi }} = \frac{{{V^2}}}{2}{\left( {1 - {A^{ - 2}}} \right)^2} - {V^2}\left( {1 - {A^{ - 2}}} \right) + \sqrt {1 + R_0^2{A^{4/3}}}  - \sqrt {1 + R_0^2}  - \frac{{{\beta ^2}{R_0}{A^{1/3}}}}{{2\pi }} + \frac{{{\beta ^2}{R_0}}}{{2\pi }}.
\end{equation}
Now, multiplying the Eq. (\ref{diff2}) by the quantity $dA/d\xi$ and finally integrating with the same boundary conditions, gives rise to the well-known energy integral of the form in terms of the plasma center of mass density variable
\begin{equation}\label{energy}
{({d_\xi } \rho)^2}/2 + U(\rho) = 0.
\end{equation}
The desired pseudopotential is then given as
\begin{equation}\label{pseudo}
\begin{array}{l}
U(\rho) = \frac{1}{{4\pi {H^2}{R_0^3}}}\left[ {4\pi {V^2}{R_0^3} - 8\pi {V^2}{R_0^3}\rho  + 3\pi R_0\sqrt {1 + {R_0^2}} \rho  - 2\pi {R_0^3}\sqrt {1 + {R_0^2}} \rho  + {R_0^4}{\beta ^2}\rho } \right. \\ - 3\pi R_0\sqrt {1 + {R_0^2}{\rho ^{2/3}}} {\rho ^{4/3}} + 4\pi {V^2}{R_0^3}{\rho ^2} + 8\pi {R_0^3}\sqrt {1 + {R_0^2}} {\rho ^2} - 4{R_0^4}{\beta ^2}{\rho ^2} \\ \left. { - 6\pi {R_0^3}{\rho ^2}\sqrt {1 + {R_0^2}{\rho ^{2/3}}}  + 3{R_0^4}{\beta ^2}{\rho ^{7/3}} - 3\pi \rho {{\sinh }^{ - 1}}R_0 + 3\pi \rho {{\sinh }^{ - 1}}\left( {R_0{\rho ^{1/3}}} \right)} \right]. \\
\end{array}
\end{equation}
The existence of nonlinear localized density structures (solitary waves) requires that the following necessary conditions to satisfy, simultaneously
\begin{equation}\label{conditions}
{\left. {U(\rho)} \right|_{\rho = 1}} = {\left. {\frac{{dU(\rho)}}{{d\rho}}} \right|_{\rho = 1}} = 0,\hspace{3mm}{\left. {\frac{{{d^2}U(\rho)}}{{d{\rho^2}}}} \right|_{\rho = 1}} < 0.
\end{equation}
A sufficient condition is to exclude the shock-like structures, i.e., for at least one either maximum or minimum there should be at least one nonzero $n$-value, with $U(\rho_{m})=0$, so that for all value of $\rho$ in the range ${\rho _m} > \rho  > 1$ (compressive soliton) or ${\rho _m} < \rho  < 1$ (rarefactive soliton), $U(\rho)$ is negative. In this way one obtains a potential minimum describing a solitary wave propagation given as
\begin{equation}\label{soliton}
\xi  - {\xi _0} =  \pm \int_1^{\rho_m} {\frac{{d\rho}}{{\sqrt { - 2U(\rho)} }}}.
\end{equation}
Obviously the two necessary conditions given in Eq. (\ref{conditions}) is met for the pseudo-potential given by Eq. (\ref{pseudo}). The third necessary condition, however, requires the direct evaluation of the second derivative of the Sagdeev potential, in Eq. (\ref{pseudo}), at the unstable point, $\rho=1$,
\begin{equation}\label{dd}
{\left. {\frac{{{d^2}U(\rho)}}{{d{\rho^2}}}} \right|_{\rho = 1}} = \frac{1}{{3{H^2}}}\left[ {6{V^2} + R_0\left( {\frac{{{\beta ^2}}}{\pi } - \frac{{2R_0}}{{\sqrt {1 + {R_0^2}} }}} \right)} \right].
\end{equation}
This puts an upper limit on soliton Mach-number for existence of a magnetoacoustic solitary excitations the value of which is defined as
\begin{equation}\label{consol2}
{V_{cr}} = \sqrt {\frac{{{R_0^2}}}{{3\sqrt {1 + {R_0^2}} }} - \frac{{R_0{\beta ^2}}}{{6\pi }}}.
\end{equation}
The critical Mach-number value obtained above, for $\beta=0$, appropriately reduces to the value given for the simple unmagnetized Fermi-Dirac plasma \cite{akbari7}. To this end, we examine the sufficient condition, i.e., the existence of the $\rho_m$ values. Thus, we examin the limits of the pseudopotential given in Eq. (\ref{pseudo}), i.e.
\begin{equation}\label{nm}
\mathop {\lim }\limits_{\rho \to 0} U(\rho) = \frac{{{V^2}}}{{{H^2}}} > 0,\hspace{3mm}\mathop {\lim }\limits_{\rho \to \infty } U(\rho) = {\mathop{\rm sgn}} ({\beta ^2} - 2\pi )\infty.
\end{equation}
It is evident that, the solitary waves can only be of rarefactive-type. The possibility of such structures has also been reported for relativistic paramagnetic Fermi-Dirac plasma \cite{akbari6}. It is also observed that above the critical value of $\beta_{cr}=\sqrt{2\pi}$ nonlinear excitations (periodic or solitary) are forbidden. In the next section we will again encounter this critical value and its prominent definition in the context of quantum collapse in orbital-ferromagnetic plasma.

\section{Discussion}\label{discussion}

Figure 1 depicts the thick curve where the effective quantum potential vanishes, i.e., $\Psi_{eq}=0$, and the filled region, where, the magnetosonic solitary excitations are stable. It is evident from Eq. (\ref{nm}) that no solitary or periodic excitations exist above a critical value of $\beta_{cr}>\sqrt{2\pi}$. It is also noted that the quantum collapse in LOFER plasma state resides in the region where the nonlinear excitations disappear. The quantum collapse occurs for $\beta  = \sqrt {2\pi /R} {(1 + {R^2})^{1/4}}$, or equivalently, for the induced field strength of $B = c\sqrt {2\pi {m_e}/{m_i}} {R^{3/2}}{(1 + {R^2})^{1/4}}$. It is noted that the $\beta_{cr}=\sqrt{2\pi}$ value corresponds to the value of $\alpha\simeq 1.754\times10^9$, which is slightly higher than the value reported by O'Connell and Roussel \cite{con3} but lower than the value reported by Pohl and Schmidt-Burgk \cite{burk1}. The vertical line in Fig. 1 corresponds to Chandrasekhar critical relativity parameter value of $R=\sqrt{2}$ which corresponds to the magnetic field strength of $B\simeq 3.882\times10^9G$ in the collapsing LOFER. The corresponding magnetic field strength can be found in a highly magnetized white dwarf. For a typical inertial confined dense plasma with electron density of $n_e\simeq10^{23}cm^{-3}$ it is easily found that, in laboratory LOFER, the quantum collapse can occur for the magnetic field of $B=9.737\times10^6G$ which might be realized in future. As it was mentioned earlier, parameter $\alpha$ and consequently $\beta$ contains a rapidly oscillating part changing as the successive Landau magnetic-levels are filled/unfiiled by the electrons, so that there is a possibility for the LOFER state to oscillate back and forth between nonlinear excitation region and quantum collapse region, when the $\beta$ is around the critical-value. This effect may lead to the magnetosonic pulsation when the density is very high or when the the Landau levels are very close to each other.

In Fig. 2 we show the volume in $V$-$R_0$-$\beta$ space where the localized nonlinear excitations exist. It is clearly remarked that, for a fixed value of $\beta$, there always exists a lower cut-off degeneracy parameter value $R_{0m}$ below which the solitary excitations are absent. It is also observed that, as the value of $\beta$ increases lower Mach-numbers will be available to localized nonlinear excitations and conversely for the fixed values of $\beta$ as the degeneracy parameter $R_0$ increases the higher Mach-numbers will be available to solitary excitations. On the other hand, the pseudopotential potential and their shape variations (when only one fractional plasma parameter is changed and others are fixed) are shown in Fig. 3. It is revealed that, the change in the pseudopotential profile with change in plasma parameters is more significant for non-relativistic degeneracy cases (left column) rather than the relativistic ones (right column). This feature is further revealed in Fig. 4, where the rarefactive localized excitation are depicted. From Fig 4. (which borrows the same parameters from Fig. 3) is easily concluded that there is fundamental differences between non-relativistic and relativistic localized nonlinear excitations. Fig. 4(a) shows that for the non-relativistic LOFER plasma the soliton amplitude/width increases/decreases as the degeneracy parameter is increased for fixed value of $\beta$, while, Fig. 4(b) reveals that in the relativistic case the amplitude is almost unaffected by this variation and only the soliton with increases with increase in the value of relativistic degeneracy. Concerning the variation of $\beta$ (or equivalently the value of induced magnetic field) one observes from Fig. 4(c) that the increase in the value of $\beta$ for a fixed value of the degeneracy parameter, $R_0$, opposite to the case of Fig. 4(a) causes a decrease/increases in the soliton amplitude/width. Finally, Fig. 4(d) reveals that the change in the parameter $\beta$ for a fixed value of degeneracy parameter value in the case of relativistic degeneracy leaves the soliton amplitude and width almost invariant opposite to the case of non-relativistic degeneracy. The mentioned fundamental differences between the cases of non-relativistic and relativistical LOFER plasma degeneracy mark distinct features similar to the ones presented in Ref. \cite{akbari6} and can help understand the underlying physical nonlinear processes in stellar dense plasmas.

\section{Summary}\label{conclusion}

We investigated the possibility of a quantum collapse and nonlinear localized excitations in a Landau orbital ferromagnetic dense stellar plasma in the framework of quantum magnetohydrodynamic model using the concepts of effective potential and Sagdeev pseudopotential. Our findings confirm that a quantum collapse is possible for magnetic field strengths present in some white dwarf stars and the value of induced field found here are consistent with the ones reported earlier. Our investigation also reveals that the magnetosonic nonlinear propagations behave rather differently in the two distinct non-relativistic and relativistic degeneracy regimes. Current findings can help better understand the origin of the most important mechanisms such as collapse and high magnetic field observed in some compact astrophysical objects.

\newpage

\textbf{FIGURE CAPTIONS}

\bigskip

Figure-1

\bigskip

The zero effective potential (quantum collapse) curve (thick-curve) and the region (filled-region), where the magnetosonic localized excitations occur for Landau orbital ferromagnetic (LOFER) plasma. The vertical line denotes the Chandrasekhar critical relativity parameter value ($R=\sqrt{2}$).

\bigskip

Figure-2

\bigskip

Figure 3 shows a volume in $V$-$\beta$-$R_0$ space in which a solitary excitation exists.

\bigskip

Figure-3

\bigskip

(Color online) The variations of pseudopotential depth and width for rarefactive solitary nonlinear magnetosonic waves with respect to change in each of three independent plasma fractional parameter, normalized soliton-speed, $V$, the ferromagnetic parameter, $\beta$, and the relativistic degeneracy parameter, $R_0$, while the other three parameters are fixed. The dash-size of the potential curves increase with regard to increase in the varied parameter.

\bigskip

Figure-4

\bigskip

(Color online) The variations of localized magnetoacoustic density-excitation profiles with respect to change in each of three independent plasma fractional parameter, normalized soliton-speed, $V$, the ferromagnetic parameter, $\beta$, and the relativistic degeneracy parameter, $R_0$, while the other three parameters are fixed. The thickness of the profile increase with regard to increase in the varied parameter.

\bigskip

\newpage

\begin{figure}[ptb]\label{Figure1}
\includegraphics[scale=.6]{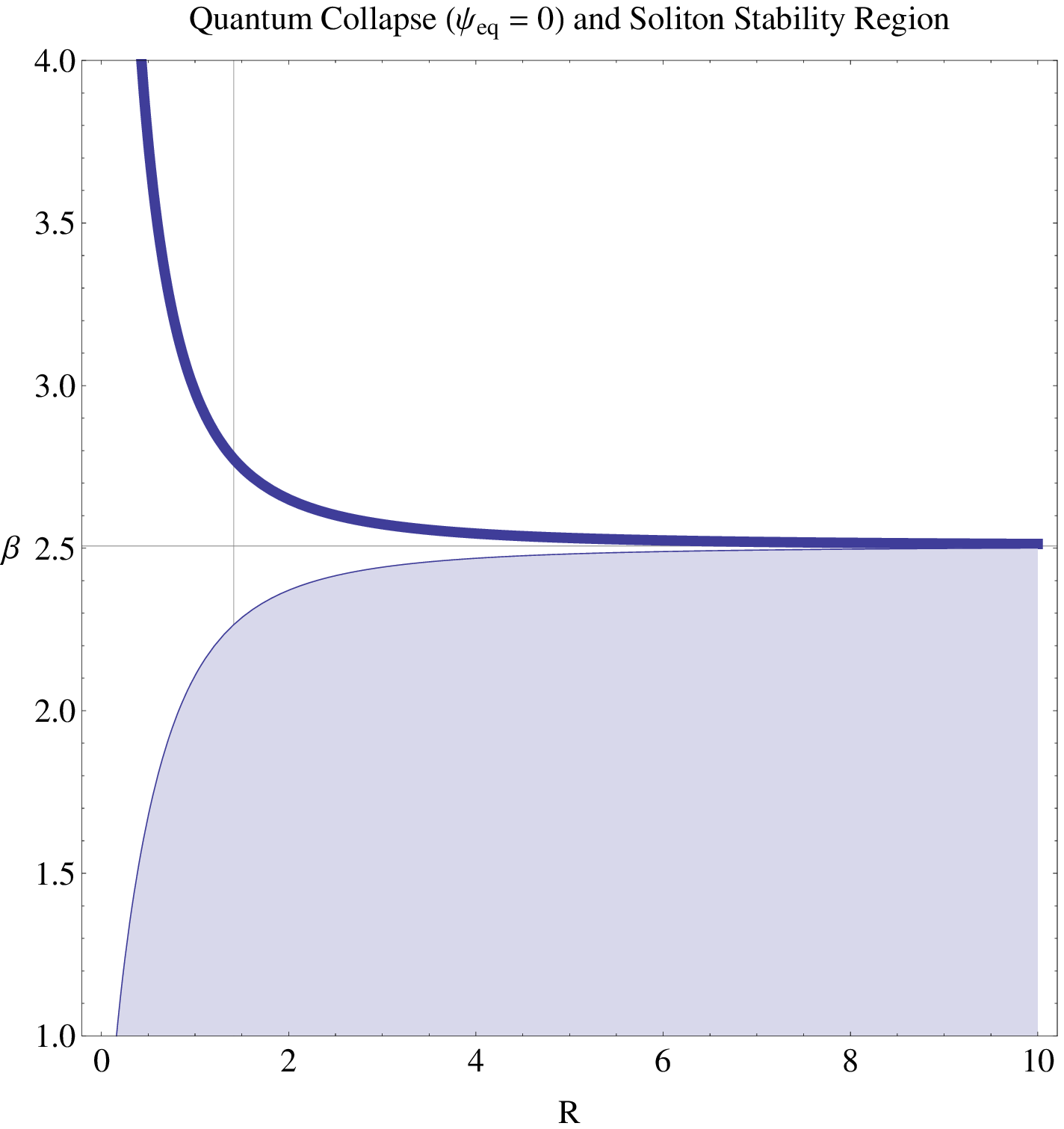}\caption{}
\end{figure}

\newpage

\begin{figure}[ptb]\label{Figure2}
\includegraphics[scale=.7]{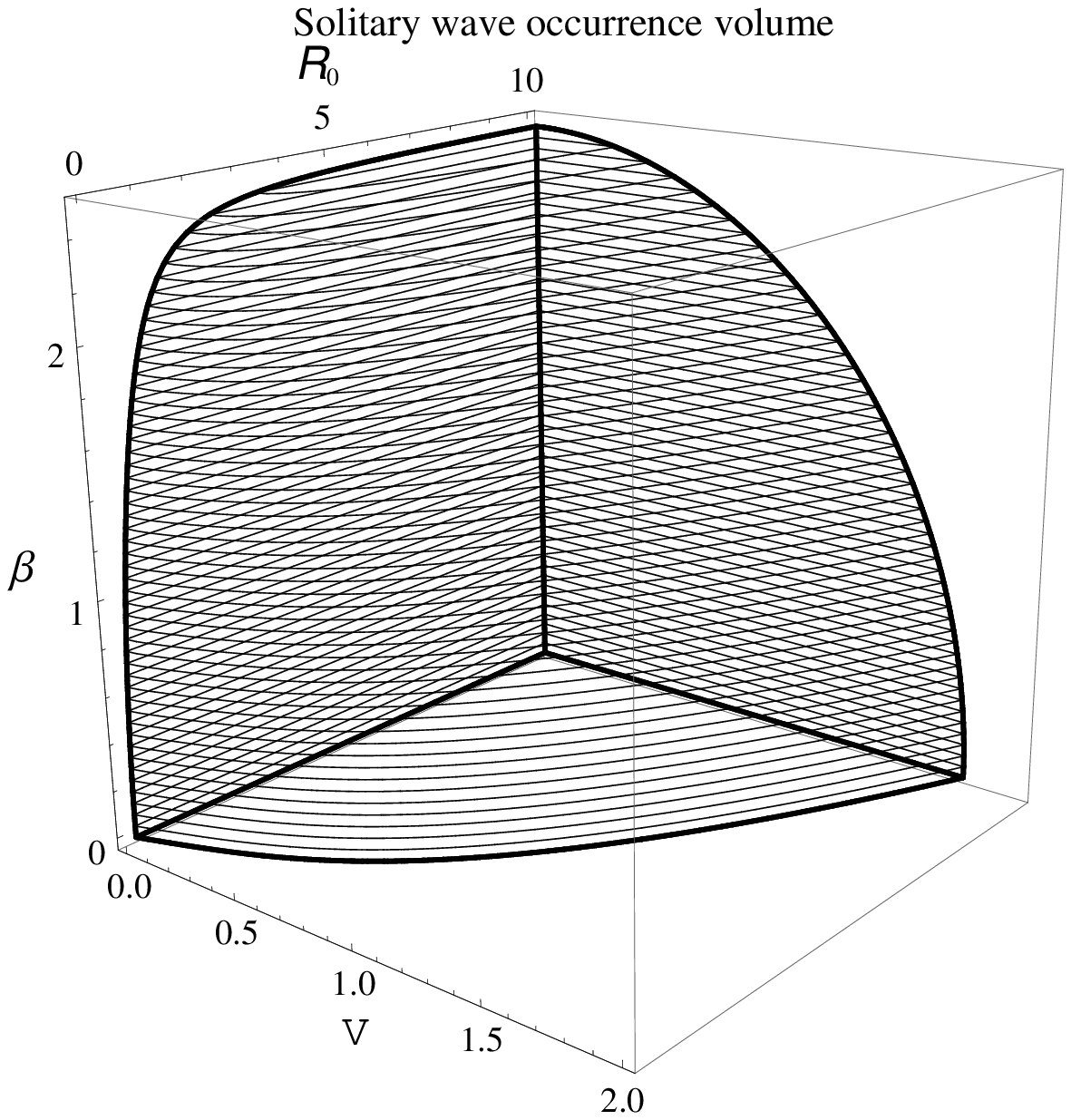}\caption{}
\end{figure}

\newpage

\begin{figure}[ptb]\label{Figure3}
\includegraphics[scale=.6]{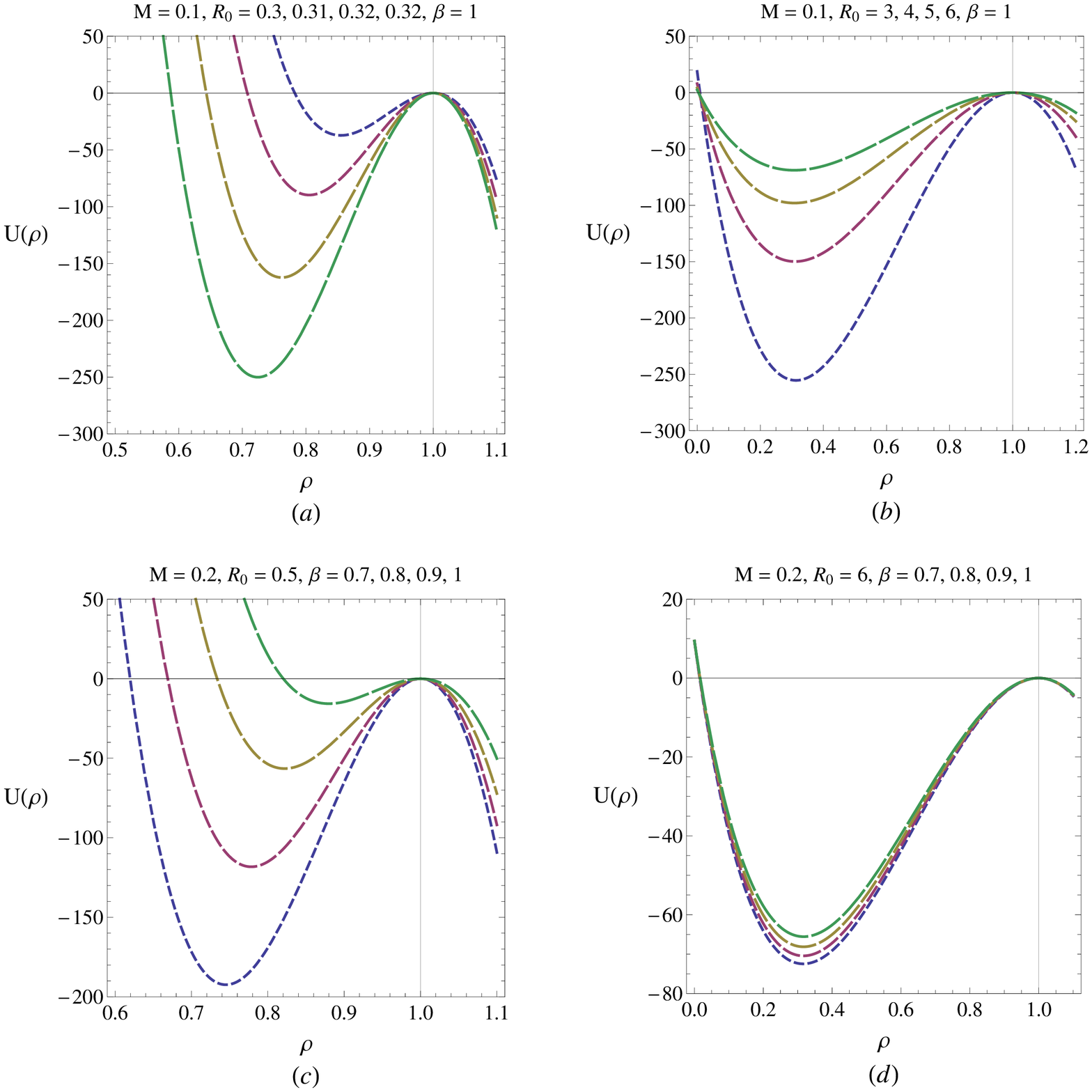}\caption{}
\end{figure}

\newpage

\begin{figure}[ptb]\label{Figure4}
\includegraphics[scale=.6]{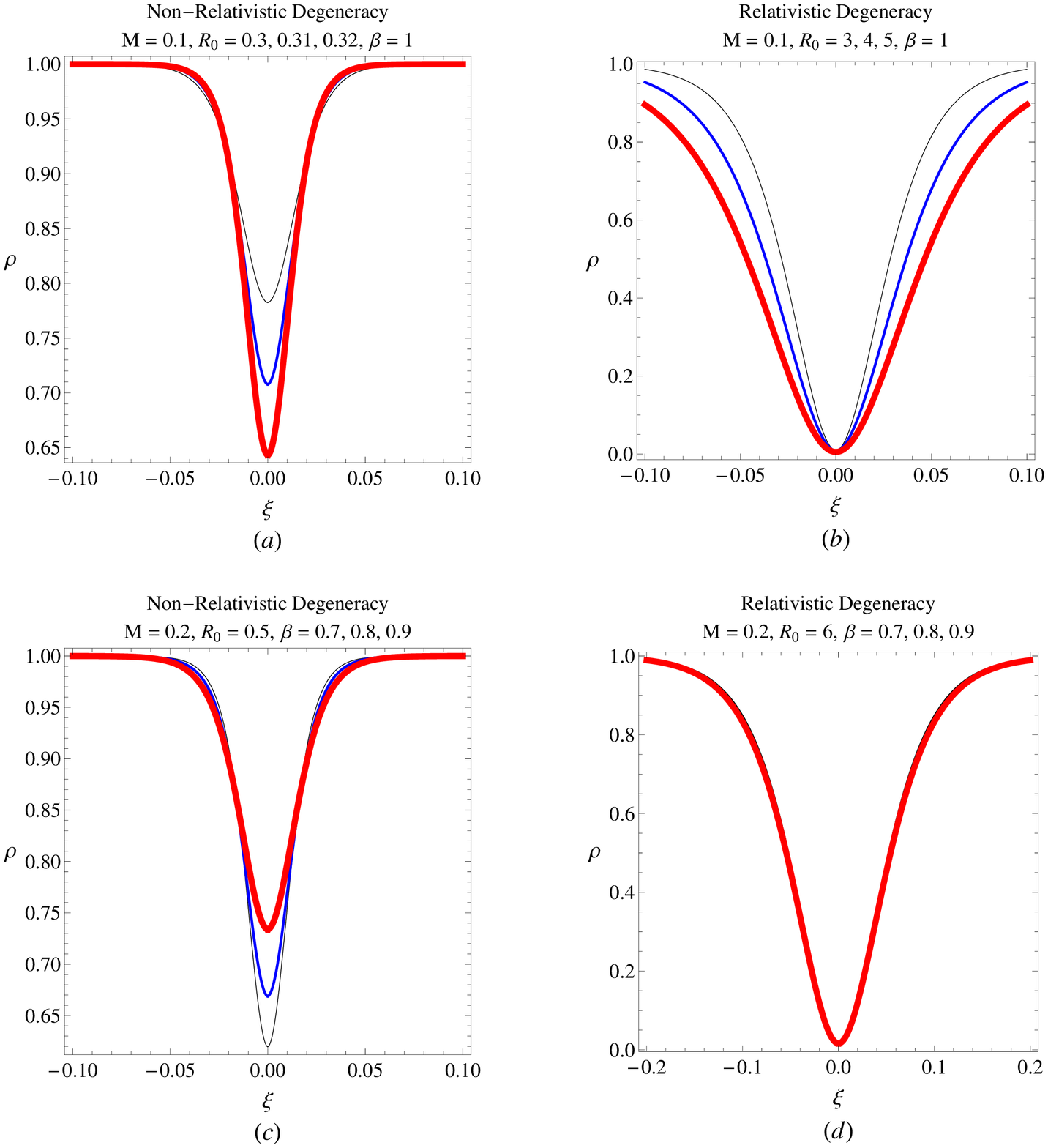}\caption{}
\end{figure}

\end{document}